\documentstyle[12pt]{article}
\begin{document}

\begin{titlepage}
~\vspace{2.5cm}

\vskip2cm
\hfill{IC/IR/99/16}
\begin{center}
\vskip2cm
\Large\bf Convergence of the expansion  of the Laplace-Borel integral in 
perturbative QCD improved by conformal mapping  
\end{center}

\vspace{1.0cm}

\begin{center}
Irinel Caprini\\
National Institute of Physics and Nuclear Engineering, POB MG 6, Bucharest,\\
R-76900 Romania\\ and\\
 The Abdus Salam International Centre for Theoretical Physics, Trieste, 
Italy \end{center}

\centerline{and}

\begin{center}
Jan Fischer\\
Institute of Physics, Academy of Sciences of the Czech Republic, 
182 21  Prague 8, \\ Czech Republic
\end{center}
\vspace{1.2cm}
\begin{abstract} The
  optimal conformal mapping of the Borel plane was recently used to  
accelerate the convergence of the perturbation expansions in QCD.  In this
work we discuss the relevance of the method for the 
 calculation of the Laplace-Borel integral expressing formally the QCD Green
functions. We define an optimal expansion of the Laplace-Borel integral in the
principal value prescription and establish  conditions under which the
expansion  is convergent. \end{abstract}
 \end{titlepage}
 \newpage
\section{Introduction} A suitable method for accelerating the convergence of 
power series is based on  conformal mappings. As is known, a power series
converges inside the circle  passing through the nearest singularity of the
function to be approximated.  Some time ago, in Ref. \cite{CiFi}, it was
shown that, if the position of the
 singularities of the expanded function is
known, one can  reach the fastest convergence rate by expanding in powers of
 the function that conformally maps the {\it whole} holomorphy domain onto a
 unit disk. In addition, the convergence region extends over the whole
holomorphy  domain. 
In a recent paper \cite{CaFi} we applied the technique proposed in  \cite{CiFi}
to the Borel transform of the Green functions in perturbative QCD. As discussed 
in \cite{CaFi}, the Borel plane is very suitable for applying the method, since 
some information about the singularities of the Borel transform 
is available from the study of certain classes of Feynman digrams and from 
nonperturbative arguments. By the technique of conformal mapping, this 
additional information can to a certain extent be incorporated even into the 
lowest-order terms. In this way, the convergence of the perturbative expansion 
is improved, allowing one in particular to approximately predict the next-order 
perturbative terms from the calculated low-order ones \cite{CaFi}.

 In Ref.\cite{CaFi} the expansion in powers of the optimal conformal mapping 
 variable\footnote{The conformal mapping that maps the whole holomorphy domain
of the expanded function onto the unit disk will be called optimal. In this
case, the singularities are mapped onto the boundary circle, and the requirement 
of holomorphy implies convergence of the power series at every point of the 
disk, which is the map of the holomorphy domain.} was also used to 
calculate the Borel-Laplace integral, which is supposed to give, with a certain 
prescription of treating the infrared renormalons, the Borel summation of the 
large orders in the Green functions. The numerical results on mathematical models 
discussed in \cite{CaFi} indicate that the power expansion in the optimal 
variable makes also the calculation of  the Borel integral convergent, in 
addition with a very high convergence rate. However, only 
 qualitative arguments explaining the results were given, and the problem 
 whether the improved  expansion of the integral is convergent, or signs of 
 divergence might appear at large orders, remained open. In the present paper 
 we address this problem and investigate the convergence of the expansion of 
 the Borel integral in perturbative QCD, improved  by the use of conformal 
 mapping.
\section{Optimal expansion of the Laplace-Borel integral}
We study the following integral
\begin{equation}\label{Laplace}
   I(a) =
    \int\limits_0^\infty\, {\rm e}^{-{u\over a}} B(u)\,{\rm d}u \,,
\end{equation}
where $B(u)$ is assumed to be analytic near $u=0$, where it can be expanded
 as a Taylor series
\begin{equation}\label{Buseries}   B(u) = \sum_{n=0}^\infty\,b_n\,u^n \,
\end{equation}
 converging inside a circle of non-vanishing radius.
The function $I(a)$ 
is of interest for the Borel summation of the Green functions in perturbative 
QCD; note, however, that the integral on the right hand side of (\ref{Laplace}) 
is ill-defined if $B(u)$ has singularities along the positive real semiaxis, 
which is the case of QCD (infrared renormalons). First, the singularities of 
$B(u)$ (renormalons of either kind) make the expansion (\ref{Buseries}) badly 
divergent along the integration path. Second, the function $B(u)$ itself is, 
because of infrared renormalons, 
not uniquely defined along 
the integration path. We shall discuss both these problems below in this paper. 

We shall consider for illustration
 the Adler function $D(s)$ of the massless QCD vacuum polarization,
which  can be expressed formally as \cite{Max}-\cite{Neub}
\begin{equation}\label{Adlerb}
   D(s) = 1 + {1\over\pi \beta_0}\,I(a)\,,
\end{equation}
with $a=\beta_{0} \alpha_{s}(-s)$, where $\alpha_s(-s)$ is the running
coupling,  and $\beta_0=(33-2 n_f)/12\pi$ is the first coefficient of the
$\beta$ function. The expression (\ref{Adlerb}) formally reproduces the  
renormalization-group-improved expansion of the Adler function
\begin{equation}\label{Dseries}
   D(s) = 1 + \sum_{n=1}^\infty\,D_n\,\bigg( {\alpha_s(-s)\over\pi}
   \bigg)^n \,,
\end{equation}
by taking the coefficients $b_n$  in the expansion (\ref{Buseries}) of the form
\begin{equation}\label{bn}
b_n= {1\over n! }\,{D_{n+1}\over (\pi \beta_0)^{n}} \,.
\end{equation}
We consider also minkowskian quantities, like  the hadronic decay rate
 of the $\tau$ lepton, $R_\tau$, which can be expressed formally as \cite{Alta}
 \begin{equation}\label{Rtaub}
R_\tau\,=\,3 (1+\delta_{{\rm EW}})\left[1+{1\over \pi \beta_0}\int_0
^\infty{\rm d}u \exp\bigg(-{u\over \beta_0\alpha_s(m_\tau^2)}\bigg)\,B(u)\,
F(u)\,\right]\,. \end{equation}
Here $\delta_{EW}$ is an electroweak correction, $B(u)$ is the Borel transform 
of the Adler function and
\begin{equation}\label{F}
F(u)={-12\,\sin(\pi u)\over \pi\,u(u-1)(u-3)(u-4)}\,.
\end{equation}
The extra factor $\sin(\pi u)$ in the Laplace-Borel integral is generic for
minkowskian  quantities. Note that strictly speaking the expressions
(\ref{Adlerb}) and  (\ref{Rtaub}) are not equivalent to the Borel summation
method, which requires  an analytic continuation of $B(u)$ from the
convergence disk to an infinite  strip of non-vanishing width, bisected by the
real positive semiaxis. This  condition is not fulfilled in QCD because of
infrared renormalons, which  produce cuts of $B(u)$ located along the real
positive semiaxis. 

We shall be concerned with the evaluation of the integral (\ref{Laplace}) for 
complex $a$ of the general form $a=|a|{\rm e}^{i\psi}$, where $\psi = 
{\rm arg}\, a$ is the phase of $a$. In the case of the Adler function, with 
the running coupling at one loop in the $V$ scheme 
$\alpha_s^{(V)}(-s)=1/[\beta_0\ln(-s/\Lambda_V^2)]$, and
writing  $-s=|s|\,e^{i(\phi-\pi)}$, we have
\begin{equation}\label{a}
{1\over a}= \ln {|s|\over \Lambda_V^2}-i(\pi-\phi)\,.
\end{equation}
Outside the Landau region, i.e. for $|s|>\Lambda_V^2$, we have $\cos \psi >0$, 
so that
\begin{equation}\label{cond}
|\psi| <{\pi\over 2}\,,
\end{equation} and $\psi$ is related to the momentum plane variable $s$ by
\begin{equation}\label{psiD}
\psi= {\rm Arctg} \left[(\pi-\phi) /\ln {|s|\over \Lambda_V^2}\right]\,.
\end{equation}
 The phase $\psi$ is positive for $s$ in the upper half of the $s$- plane, 
 where $0<\phi<\pi$,  negative for $s$ in the lower half-plane, where 
 $\pi<\phi<2\pi$,  and $0$ along the euclidean axis.

For the  minkowskian quantity (\ref{Rtaub}) we combine the additional factors 
$\exp (\pm i \pi u)$ due to the sinus with the exponential, which amounts to 
taking $a$ complex with 
\begin{equation}\label{psiR}
 \psi=\pm{\rm Arctg}\,[\pi \bar a]\,,
\end{equation}
where $\bar a=\beta_0 \alpha_s(m_\tau^2)$.

As already mentioned, the Borel transform $B(u)$ has singularities in the 
complex plane, correlated to the factorial increase of the perturbative 
coefficients of the Green functions  at large orders \cite{Muel}, \cite{Broa}.
The precise form of the singularities is not known for the exact theory, but  
the position and the nature of the first renormalons can be inferred from 
general principles. In the case of the Adler function the first ultraviolet 
(UV) renormalon is situated at $u=-1$ and the first infrared (IR) one at 
$u=2$, and they are branch points of the type $(1+u)^{\gamma_1}$ and  
$(2-u)^{\gamma_2}$ respectively, with $\gamma_1$ computed in Ref. \cite{BBK} 
and $\gamma_2$ in \cite{Muel}. We mention also  that the  summation of the 
one-renormalon chains in  massless QCD in the large $\beta_0$ limit gives 
\cite{Broa}, \cite{Bene}:
\begin{equation}\label{largeb}
    B(u) = \frac{32 {\rm e}^{-Cu}}{3(2-u)}\,\sum_{k=2}^\infty\,
   \frac{(-1)^k\,k}{\big[ k^2-(1-u)^2 \big]^2} \,,
\end{equation}
i.e. all the singularities are poles ($C$ is a scheme-dependent constant, with 
$C=-5/3$ in the $\overline{\rm MS}$ scheme, and $C=0$ in the V scheme)  
\cite{Max}, \cite{Neub}. 

The series ({\ref{Buseries}) converges only inside the circle $\vert u\vert 
<R$ passing through the nearest singularity ($R=1$ for the Adler function). 
Since the integration range in (\ref{Laplace}) extends far outside this  
region, by inserting ({\ref{Buseries}) in 
(\ref{Laplace}) and integrating term by term one obtains a divergent expansion. 
By the technique of conformal mappings one extends the 
domain of convergence of a series beyond the limit imposed by the first 
singularity.  In \cite{CaFi} we used the optimal conformal mapping 
\begin{equation}\label{wborel}
w=w(u)={\sqrt{1+u}-\sqrt{1-u/2}\over \sqrt{1+u}+\sqrt{1-u/2}}\,,
\end{equation} with  the inverse 
$u(w)=8w/(3-2w+3w^2)$.
The transformation (\ref{wborel}) preserves the origin and maps the complex  
$u$ plane, cut along the real axis for $u>2$ and for $u<-1$, onto the interior
of the circle $\vert w\vert\, <\, 1$, all the singularities of the Borel 
transform  being mapped onto the boundary  $\vert w\vert\, =\, 1$.
The expansion in powers of $w$ , 
\begin{equation}\label{Bwseries}
B(u)=\sum_{n=0}^\infty c_n\,w^n\,,
\end{equation}
is called optimal because it converges inside the circle $\vert w\vert <1$, 
i.e. {\it in the whole domain of holomorphy of} $B(u)$ (which is the doubly cut
complex $u$-plane in our case), up to points close to the  branch cuts produced 
by renormalons. As was already pointed out above, this power 
expansion yields, when compared with other conformal mappings, {\it the 
fastest large-order convergence rate} (see a proof in \cite{CiFi}).
In practice, as discussed in \cite{CaFi}, the expansion (\ref{Bwseries}) is
obtained by suitably reorganizing the summation of the original series
(\ref{Buseries}). More precisely, consider the expansion of each $u^n$ in
powers of $w$, truncated at a finite order $N$. In particular, in our case this
expansion has the general form \begin{equation}\label{uNn} u_{N}^n =
\sum_{j=n}^{N} c_{nj} w^j \,, \end{equation}
with the coefficients $c_{nj}$ obtained by expressing $u$ in terms of $w$
(in our case $u(w)$ is given explicitly  after formula 
(\ref{wborel})). Starting now with
the expansion (\ref{Buseries}) truncated at finite order $N$, and replacing
each  $u^n$ by its approximant $u_{N}^n$, one
obtains a truncated expansion of the function $B$ in powers of $w$, 
which in the limit $N\to \infty$ gives (\ref{Bwseries}).

  By inserting
the optimal expansion (\ref{Bwseries}) into the integral (\ref{Laplace}) we 
obtain the formal development \begin{equation}\label{improv}
I(a)=\sum\limits_{n=0}^\infty c_n I_n(a)\,, \end{equation}
with 
\begin{equation}\label{In}
I_n(a)=\int\limits_0^\infty {\rm e}^{-{u\over a}} w^n\,{\rm d} u\,.
\end{equation}
In the present work we shall adopt (\ref{improv}) as the optimal expansion 
of the Laplace-Borel integral. We point out that in the physical case this 
seems to be a natural definition.  Indeed, when attempting to
make the Borel summation of a perturbation expansion in QCD, one starts
with a finite sum of the form
\begin{equation}\label{inprob}
I_{N}(a)=\sum\limits_{n=0}^{N} b_n \int\limits_0^\infty 
{\rm e}^{-{u\over a}} u^n\,{\rm d} u\,.
\end{equation}
By replacing here the powers $u^n$ with the approximations (\ref{uNn}),
 we replace $I_{N}(a)$ by an expansion of the form
\begin{equation}\label{improvN}
\sum\limits_{n=0}^{N} c_n I_n(a)\,,
\end{equation}
which in the limit $N \rightarrow \infty$ leads to (\ref{improv}). 

Actually, as mentioned above, the conditions for the Borel summation are,
because of the infrared renormalons, not fulfilled. Therefore, Eq.
(\ref{improv}) can be considered as a definition of $I(a)$, provided that (i)
the integration path in expressions like  (\ref{Laplace}) or (\ref{In})
 is consistently defined, and (ii) the series (\ref{improv}) is convergent.
Let us devote a brief discussion to these conditions. 

(i) As concerns the integration contour, let us notice that the
expansion (\ref{improv}) has not a precise mathematical sense with the 
$I_{n}(a)$ defined by (\ref{In}), because the integration path runs along the 
positive real semiaxis, where the $w^{n}$ have cuts. We shall adopt, as in 
\cite{CaFi}, the generalized principal value (PV) prescription, 
defining the $I_{n}^{PV}(a)$ as
\begin{equation}\label{Inpm}
I_{n}^{PV}(a)={1\over 2}\int\limits_{\cal C_+}\, {\rm e}^{-{u\over a}}\, (w(u))^n
\,{\rm d}u +{1\over 2} \int\limits_{\cal C_-}\, {\rm e}^{-{u\over a}}\, (w(u))^n
\,{\rm d}u  \, 
\end{equation}
for $n=0,1,2,...$, where ${\cal C_+}$ (${\cal C_-}$) are lines parallel to the 
real positive axis, slightly above (below) it.  
While the PV 
prescription does not always give the expected results \cite{Zinn}, in QCD it 
has the advantage that it reproduces, to a larger extent than other choices, 
the momentum plane analyticity properties of the Green functions derived from 
the general principles of field theory. In particular, as discussed in 
\cite{CaNe}, the Adler function calculated with the PV prescription has no 
unphysical singularities in the region $|s|> \Lambda^2$. 
The functions $I^{PV}_{n}(a)$, $n=1,2...$ are chosen so as to share some of
the known properties with the unknown $I^{PV}(a)$. This makes them suited
for the definition of $I^{PV}(a)$ by means of the expansion      
\begin{equation}\label{improvPV}
I^{PV}(a)=\sum\limits_{n=0}^\infty c_n I^{PV}_n(a)\,.
\end{equation}

(ii) The convergence of the series (\ref{improvPV}) for complex $a$ is not a
priori  obvious. Indeed, the expansion (\ref{Bwseries}) converges at points
$|w|<1$,  therefore in the neigbourhood of the integration axis, but not 
necessarily 
on the boundary. 
 One might therefore expect that the boundary singularities could manifest in a 
 dramatic way for very large orders $N$, making the series (\ref{improvPV}) 
 divergent, like in the case of the original expansion (\ref{Buseries}). In 
\cite{CaFi} we investigated mathematical models with $B(u)$ having a few number 
of isolated branch point singularities, and real values of $a$. The numerical 
results confirm that the expansion (\ref{Buseries}) in powers of $u$ gives 
results which deviate  dramatically from the exact value for large $N$, which 
is typical for a divergent  
expansion. On the other hand, the improved series (\ref{Bwseries}) 
in powers of the optimal variable led to results improving continuously with 
increasing $N$, and no signs of divergence appeared even at very high $N$. In 
the next Section we shall discuss the convergence of the optimal
expansion of the Laplace-Borel integral, bringing analytic arguments which
explain the numerical results obtained in  \cite{CaFi}. 

\section{Convergence of the optimal expansion}
We investigate the expansion (\ref{improvPV}) with the functions 
$I^{PV}_{n}(a)$ defined by means of the PV prescription (\ref{Inpm}). 
We consider in our discussion analytic functions $B$ 
of  real type, i.e. which satisfy $B^*(u)=B(u^*)$, where $u^*$ is the complex  
conjugate of $u$. Therefore, the coefficients $b_n$ in the expansion  
(\ref{Buseries}), as well as the coefficients $c_n$ in the expansion  
(\ref{Bwseries}) are real. 

The contribution to (\ref{Inpm}) of
integral along the contour ${\cal C}_+$ can be written as
 \begin{equation}\label{Inp}
I^+_n(a)=\int\limits_{{\cal C}_+} {\rm e}^{-F_n(u)}\,{\rm d} u \,,
\end{equation}
  where
\begin{equation}\label{Fn}
 F_n(u)={u\over a}- n \ln w(u)\,.\end{equation}
We evaluate the integral (\ref{Inp}) for large $n$ by applying the method of 
steepest descent \cite{Jeff}, \cite{Zinn}. The saddle points are given by  the 
equation
\begin{equation}\label{eq}
{w'(u)\over w(u)}={1\over a n}\,,
\end{equation} 
which has four solutions, having at large $n$ the form
\begin{equation}\label{saddle}
 {1+i\over 2^{1/4}}  \sqrt{a n}\,,\,\,\, { 1-i\over 2^{1/4}}  \sqrt{a n}\,,\,\,\,
{-1+i\over 2^{1/4}} \sqrt{a n}\,,\,\,\,{-1-i\over  2^{1/4}} \sqrt{a n}\,.
\end{equation} 
Of interest for the evaluation of (\ref{Inp}) is the point 
\begin{equation}\label{u0p}
u_0= 2^{-1/4} ( 1+i) \sqrt{a n}=|u_0| {\rm e}^{i\alpha}\end{equation}
with
\begin{equation}\label{u0p1}
|u_0|= 2^{1/4} \sqrt{|a| n}\,,\quad \alpha={\pi\over 4}+{\psi\over 2}\,,
\end{equation}
which is situated in the first quadrant of the $u$-plane. Indeed, since the 
phase $\psi$ of the parameter $a$ satisfies the condition (\ref{cond}), then 
${\rm Re} u_0>0$ and ${\rm Im} u_0>0$.  

 Near the saddle point $F_n(u)$ can be expanded as
\begin{equation}\label{steep}
F_n(u)=F_n(u_0)+{1\over 2}F''_n(u_0) (u-u_0)^2+...\,.
\end{equation} 
By using the expansion of $w(u)$ for large $u$ in the upper half plane 
( $w\approx \zeta (1-i\sqrt{2}/u)$, where $\zeta =(\sqrt{2}+i)/(\sqrt{2}-i)$), 
 we obtain after a straightfoward calculation
\begin{equation}\label{Fn0}
{\rm e}^{-F_n(u_0)}\approx \zeta^n\, \left(1-{2^{3/4}i\over (1+i)\sqrt{a n}}
\right)^n {\rm e}^{-2^{-1/4}(1+i) \sqrt{{n\over a}}}\approx  \zeta^n\
{\rm e}^{-2^{3/4}(1+i) \sqrt{{n\over a}}} 
\end{equation} 
and 
\begin{equation}\label{F2n0}
F_n''(u_0)\approx {2^{1/4}(1-i)\over \sqrt{n a^3}}\,= |F_n''(u_0)|{\rm e}^{i \beta}\,,
\end{equation}
where
$$ |F_n''(u_0)|={2^{3/4}\over\sqrt{n |a|^3}}\,,\quad 
\beta=-{\pi\over 4}-{3\psi\over 2}\,.$$
 Therefore (\ref{Inp}) becomes
\begin{equation}\label{Inp1}
I^+_n(a)\approx \zeta ^n  {\rm e}^{-2^{3/4}(1+i) \sqrt{{n\over a}}}\int\limits_
{{\cal C}_+}
 {\rm e}^{-{ |F_n''(u_0)| \over 2}\,{\rm e}^{i\beta}\,(u-u_0)^2}\,{\rm d} u\,.
\end{equation} 
In order to evaluate the integral we first rotate the contour ${\cal C}_+$ in 
the trigonometric direction in the upper half-plane, until it becomes a line 
passing through the origin and the saddle point $u_0$.
The rotation is possible since $B(u)$ (and therefore also $w$) has no 
singularities outside the real axis, and the arc of the circle at infinity 
gives a vanishing contribution, as can be easily verified.  Along the rotated 
line $u= {\rm e}^{i\alpha}\,t$, where $\alpha$ is the phase of $u_0$ defined in 
(\ref{u0p1}) and $t$ is real, so the integral in (\ref{Inp1}) becomes
\begin{equation}\label{Inp11}
 {\rm e}^{i\alpha}\,\int\limits_0^\infty
 {\rm e}^{-\left[{ |F_n''(u_0)| \over 2}\,{\rm e}^{i(2\alpha+\beta)}\,(t-|u_0|)^2\right]}{\rm d} t\,.
\end{equation} 
Since $\cos (2\alpha+\beta)>0$ for $\psi$ satisfying the condition ({\ref{cond}), 
the integration axis lies in the two valleys near the saddle point $u_0$. 
Therefore it can be deformed into the path of steepest descent through $u_0$, 
without passing outside the valleys. We take the integral along the path going 
to infinity 
\begin{equation}\label{steepline}
u- u_0 \approx \sqrt{2 /|F_n''(u_0)|} \,{\rm e}^{-i\beta/2}\,\rho\,
\end{equation}
with real $\rho$. The phase of $(u-u_0)^2$ exactly compensates the phase
of $F''_n(u_0)$, making
 the exponent of the integrand in (\ref{Inp1}) real. The integrand  can be 
 written as ${\rm e}^{-\rho^2}$ and the integral done explicitly gives
\begin{equation}\label{Inp2}
I^+_n(a)\approx \zeta ^n  {\rm e}^{-2^{3/4}(1+i) \sqrt{{n\over a}}}
 {{\rm e}^{-i\beta/2}\over
 \sqrt{ |F_n''(u_0)|/ 2}}  \,{\sqrt{\pi}\over 2}\,,
\end{equation} 
i.e. up to a constant independent of $n$
\begin{equation}\label{Inpsaddle}
I^+_n(a)\approx n^{{1\over 4}} \zeta^n
 {\rm e}^{-2^{3/4} (1+i) \sqrt{{n\over a}}}\,.
\end{equation}
It is important to note that the path of steepest descent must not cross the 
real axis, where $B(u)$ has singularities. From (\ref{steepline}) this implies 
$-\beta/2= \pi/8+3\psi/4>0$ which writes as
\begin{equation}\label{cond+}
\psi>-{\pi\over 6}\,.
\end{equation}  

The evaluation of the integral along the contour ${\cal C}_-$ in (\ref{Inpm})
proceeds in a  similar way. The saddle point of interest is 
\begin{equation}\label{u0m}
u'_0=2^{-1/4} (1- i) \sqrt{an}\,= \, 2^{1/4} \sqrt{|a| n}\,
{\rm e}^{-i({\pi\over 4}-{\psi\over 2})}  \,,
\end{equation}
which satisfies  ${\rm Re} u_0>0$ and ${\rm Im} u_0<0$ for $\psi$ in the range
given in (\ref{cond}). Instead of (\ref{Inp1}) we have
\begin{equation}\label{Inm1}
I^-_n(a)\approx (\zeta^*) ^n  {\rm e}^{-2^{3/4}(1-i) \sqrt{{n\over a}}}
\int\limits_{{\cal C}_-}
 {\rm e}^{-{ |F_n''(u'_0)| \over 2}\,{\rm e}^{ i\beta'}\,(u-u'_0)^2}\,{\rm d} u\,,
\end{equation} 
where $\beta'=\pi/ 4-3\psi/ 2\,$. 
We rotate the contour ${\cal C}_-$ in the lower half-plane up to a line passing 
through the  point $u'_0$, and then deform it into the steepest descent path. 
One can easily verify that this path does not cross the real axis for 
$-\pi/8+3\psi/4<0$ which writes as
\begin{equation}\label{cond-}
\psi<{\pi\over 6}\,.
\end{equation} 
Collecting the terms we obtain the coefficients $I_n(a)$ in the PV
prescription (\ref{Inpm}) as \begin{equation}\label{Insaddle}
I_n(a)\approx n^{{1\over 4}} \zeta^n 
 {\rm e}^{-2^{3/4} (1+i) \sqrt{{n\over a}}} + n^{{1\over 4}} (\zeta^*)^n
 {\rm e}^{-2^{3/4} (1- i) \sqrt{{n\over a}}}\,.
\end{equation} 
In order to examine the convergence of the expansion (\ref{improvPV}), we 
consider the ratio
\begin{equation}\label{ratio}
\bigg\vert{c_n I_n(a)\over c_{n-1} I_{n-1}(a)}\bigg\vert\,,
\end{equation}
for large $n$. If the coefficients $c_n$ do not grow too rapidly, i.e.
\begin{equation}\label{cbeh} |c_n| < C {\rm e}^{\epsilon n^{1/2}}\,,
\end{equation}
for all $\epsilon >0$,
then the expansion (\ref{improvPV}) converges for $a$ complex in the domain
\begin{equation}\label{domain}
{\rm Re}[(1\pm i) a^{-1/2}]>0\,.
\end{equation}
As we already discussed these conditions are equivalent to (\ref{cond}).
If the coefficients $c_n$ behave at large $n$ like 
\begin{equation}\label{cbeh1} |c_n| \approx  {\rm e}^{c n^{1/2}}
\end{equation}
for some positive 
$c$, then the expansion (\ref{improvPV}) converges in the domain 
\begin{equation}\label{domain1}
{\rm Re}[(1\pm i) a^{-1/2}+c ]>0\,,
\end{equation}
while for coefficients $c_n$ which grow faster than $ {\rm exp }(c n^{1/2})$
the new series (\ref{improvPV}) is also divergent. We mention that such a 
behaviour is not excluded in general for series of the form (\ref{Bwseries}) 
with a radius of convergence equal to 1 \cite{Zinn}.

 We recall however that the expression (\ref{Insaddle}) is valid only for 
$\psi$ which satisfy the conditions (\ref{cond+}) and (\ref{cond-}), i.e.
\begin{equation}\label{condpm}
|\psi|<{\pi\over 6}\,,
\end{equation} which define a sector in  the $a$- complex plane (we recall that 
$\psi$ is the phase of $a$). This inequality is a condition of applicability of 
the steepest descent method used by us. We found therefore that the 
expansion  
(\ref{improvPV}), improved by the optimal conformal mapping of the Borel
plane,  is convergent if the Taylor coefficients $c_n$ of the expansion
(\ref{Bwseries})  satisfy the condition (\ref{cbeh}), at least inside the
sector (\ref{condpm})  of the complex plane of $a$, or, if they behave like
(\ref{cbeh1}), in the  smallest of the domains (\ref{domain1}) and
(\ref{condpm}). 
For the Adler function in massless QCD, using (\ref{psiD}) we write the
condition (\ref{condpm})  in the form
\begin{equation}\label{condD}
 \vert \pi-\phi \vert <   {1\over \sqrt{3}}  {\ln {|s|\over \Lambda_V^2}}\,,
\end{equation}
where $\phi$ is the phase of $s$ and we have $|s|>  \Lambda_V^2$.
For the minkowskian quantities, from (\ref{psiR}) we obtain
\begin{equation}\label{condR}
\pi \bar a < {1\over \sqrt{3}}\,,
\end{equation}
which means in particular that for the $\tau$- hadronic decay rate the 
expansion defined as in (\ref{improvPV}) is convergent for $\alpha_s(m_\tau^2)<
4/(9\sqrt{3})  \approx 0.257$.

 The behaviour of the coefficients $c_n$ depends on the singularities of
 $B(w)$. By the conformal mapping (\ref{wborel}) all the renormalons are 
 situated on the circle $|w|=1$, appearing in conjugate  pairs since $B(u)$ is 
 of real type. Assuming that all the singularities are poles or branch points, 
 $c_n$ has the generic form
\begin{equation}\label{ck}
c_n\approx {1\over n!}\, {\rm Re}\, \sum\limits_j r_j p_j(p_j+1)(p_j+2)...(p_j+n)
{\rm e}^{i\beta_j \gamma_j (p_j+n)}\,,
\end{equation}
 where $\exp (\pm i\beta_j)$ denote the position of the renormalon in the 
 $w$-plane, $r_j$ the residue, and $p_j$ the exponent of the singularity. In 
 Ref. \cite{CaFi} we investigated simple models with a finite number of 
 singularities, and real values of the parameter $a$, for which the conditions 
 of convergence are satisfied. In the physical case, one knows only that for 
 the first UV renormalon  $\alpha_1=\pi$ and $p_1=2 \gamma_1$, and for the 
 first IR renormalon $\alpha_2=0$ and $p_2=2 \gamma_2$.  In the large $\beta_0$ 
 case, as seen from (\ref{largeb}), all the singularities are poles, $p_j$ in 
 (\ref{ck}) is independent of $j$, and $r_j$ are known. In this case the 
 condition of convergence (\ref{cbeh}) is satisfied. Therefore, the optimal 
 expansion on the Laplace-Borel integral, in the PV prescription, for the
summation   of one renormalon chains in the large $\beta_0$ limit, is
convergent, at   least in the sector of the complex $a$ plane defined by the
condition  (\ref{condpm}). 

In conclusion, we investigated the  expansion of the Laplace- Borel integral in 
perturbative QCD, improved by the analytic continuation of the Borel transform 
outside the perturbative convergence disk (and, simultaneously, by reaching the
fastest convergence rate) by means of the optimal conformal mapping 
 \cite{CaFi}. The convergence properties of the new expansion depend on the 
 strength of the singularities of the Borel transform, reflected in the 
 behaviour of the Taylor coefficients of the expansion (\ref{Bwseries}). If the 
 Taylor coefficients satisfy the condition (\ref{cbeh}), the new expansion of 
 the Laplace- Borel integral converges in the sector of the complex plane of the 
 coupling $a$ defined by (\ref{condpm}). The conditions are satisfied in the 
 case of the resummation of one-loop renormalons in the large- $\beta_0$ limit. 
We mention that in the region where the series converges the function $I(a)$ 
must be analytic. For the Adler function in the complex momentum  plane this
corresponds  to the region  described by Eq. (\ref{condD}), where $D(s)$ is
analytic. 

\vskip1cm
{\bf Acknowledgements:}  One of us (I.C.) thanks Prof. S. Randjbar-Daemi  for
his kind hospitality at the High Energy Section of the Abdus Salam International 
Centre of Theoretical Physics, Trieste. The other author (J.F.) is indebted to  
Prof. A. De R\'{u}jula for hospitality at the CERN Theory Division.

\end{document}